\documentclass[10pt,journal,final,finalsubmission,twocolumn]{IEEEtran}
\usepackage{epsfig}
\usepackage{url}
\usepackage{amsmath, amsthm, amssymb}

\newtheorem{Lemma}{Lemma}

\newtheorem{Theorem}{Theorem}

\usepackage{algorithm}
\usepackage{algorithmic}
\usepackage{multirow}
\usepackage{amstext}

\usepackage{url}
\usepackage{color}

\newcommand{\warn}[1]{}
\newcommand{\nop}[1]{}

\begin{document}


\title{DCE: A Novel Delay Correlation Measurement for Tomography with Passive Realization}
\newcommand{\superast}{\raisebox{9pt}{$\ast$}}%
\newcommand{\superdagger}{\raisebox{9pt}{$\dagger$}}
\newcommand{\superddagger}{\raisebox{9pt}{$\ddagger$}}
\newcommand{\superS}{\raisebox{9pt}{$\S$}}
\newcommand{\superP}{\raisebox{9pt}{$\P$}}

\author{\IEEEauthorblockN{Peng Qin\IEEEauthorrefmark{1}\IEEEauthorrefmark{2}, Bin Dai\IEEEauthorrefmark{1}, Kui Wu\IEEEauthorrefmark{2}, BenxiongHuang\IEEEauthorrefmark{1} and Guan Xu\IEEEauthorrefmark{1}}\\
\IEEEauthorblockA{\IEEEauthorrefmark{1}Department of Electronics and Information Engineering \\
Huazhong University of Science and Technology, Wuhan, China\\}
\IEEEauthorblockA{\IEEEauthorrefmark{2}Department of Computer Science\\
University of Victoria, Victoria, Canada\\
Email: \{qinpeng, daibin, huangbx\}@hust.edu.cn, wkui@cs.uvic.ca, guanxu86@gmail.com}
}


\maketitle

\begin{abstract}

Tomography is important for network design and routing optimization. Prior approaches require either precise time synchronization or complex cooperation. Furthermore, active tomography consumes explicit probeing resulting in limited scalability. To address the first issue we propose a novel Delay Correlation Estimation methodology named DCE with no need of synchronization and special cooperation. For the second issue we develop a passive realization mechanism merely using regular data flow without explicit bandwidth consumption. Extensive simulations in OMNeT++ are made to evaluate its accuracy where we show that DCE measured delay correlation is highly identical with the true value. Also from test result we find that mechanism of passive realization is able to achieve both regular data transmission and purpose of tomography with excellent robustness versus different background traffic and package size.
\end{abstract}


\vspace{-0.1cm}
\section{Introduction}
\label{sec::introduction}
\vspace{-0.1cm}

Network tomography \cite{Y.Var::NTESD::1996} studies internal characteristics of Internet using information derived from end nodes. One advantage is that it requires no participation from network elements other than the usual forwarding of packets while traditional \emph{traceroute} method needs response to ICMP messages facing challenge of anonymous routers \cite{R.V.B::TIAR::2003}.

\nop{Network tomography can be classified into two types which are the active tomography\cite{Rab::MulSouMulDesNT::2004}, \cite{CaceresR::MulInfNetInterLos::1999}, \cite{FraLo::MulInfNetInterDis::2002} and passive tomography\cite{JinCao::TimeVarNT::2000}, \cite{Venkata::PassNTBayInf::2002}, \cite{FabioRicciato::PasTom3GNet::2006}, \cite{PasNTomErrNetNCAppr::HYao::2012}. Though both of them are more robust than traditional \emph{traceroute} approach the latter outperforms the former since it does not send extra probing packets thus consumes less bandwidth.}

Many literatures choose delay to calculate correlation between end hosts for tomography. However, they require either precise time synchronization or complex cooperation.
Moreover, active way consumes quantities of explicit probing bandwidth which results in limited scalability.

In this paper we propose a novel Delay Correlation Estimation approach named DCE with no need of cooperation and synchronization between end nodes. The greatest property is that we only need to measure the packet arriving time at receivers. To further reduce bandwidth consumption a passive mechanism using regular data flow is developed.

We do extensive simulations in OMNeT++ to evaluate its accuracy. Results show that $\sigma_{d_a,d_b}^2$ measured by DCE is highly identical with the true value $\sigma_s^2$ on shared path. By altering background traffic and package size we see that passive mechanism has excellent robustness and is able to achieve regular data transmission as well as purpose of tomography.

\vspace{-0.03cm}
\subsection{Contributions}
\vspace{-0.01cm}

\nop{The main contributions are summarized as follows.}

\begin{itemize}
  \item     We propose DCE to estimate delay correlation. This method needs no special cooperation or synchronization and avoids issues using RTT, making it largely different from prior tomography tools \cite{YolanTsa::NetDelayTomo::2003} and \cite{YolanTsa::NetRadar::2004}.
  \item  	We develop a passive mechanism for realization that is efficient for bandwidth saving.
  \item 	Extensive simulations in OMNeT++ demonstrates its accuracy and robustness.
\end{itemize}

\vspace{-0.1cm}
\section{Related work}
\label{sec::related work}
\vspace{-0.1cm}

Y. Vardi was one of the first to study network tomography \cite{Y.Var::NTESD::1996} that can be implemented in either an active or passive way. Active network tomography \cite{Rab::MulSouMulDesNT::2004}, \cite{CaceresR::MulInfNetInterLos::1999}, \cite{FraLo::MulInfNetInterDis::2002} needs to explicitly send out probing messages to estimate the end-to-end path characteristics, while passive network tomography \cite{JinCao::TimeVarNT::2000}, \cite{Venkata::PassNTBayInf::2002}, \cite{FabioRicciato::PasTom3GNet::2006}, \cite{PasNTomErrNetNCAppr::HYao::2012} infers network topology without sending any explicit probing messages.

Article \cite{YolanTsa::NetDelayTomo::2003} describes delay tomography which however, needs synchronization and cooperation between sender and receiver. In \cite{YolanTsa::NetRadar::2004} authors develop Network Radar based on RTT trying to solve these issues. However, two reasons distort the measurement accuracy. One is due to the variable processing delay at destination nodes and the other is its violation of a significant assumption that return paths of packet are uncorrelated while actually they overlaps.

In addition, delay correlation can be further used for topology recovery \cite{BDEriksson::ToPraNTIntTopoDiscov::2010}, \cite{JianNi::EffDynaRouTopoInfEnd2End::2010} that is important to improve network performance.

\vspace{-0.1cm}
\section{DCE for Delay Correlation Measurement}
\label{sec::delay correlation}
\vspace{-0.1cm}

\begin{figure}[thb]
\begin{center}
\epsfig{file=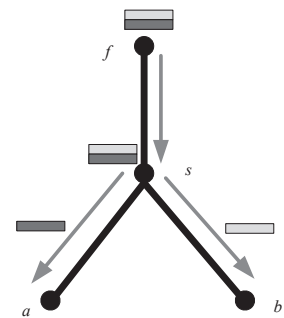, width=0.2\textwidth}
\vspace{-4mm}
\end{center}
\caption{\label{fig::network_model} The tree structure: a sender $f$ and two receivers $a$, $b$}
\vspace{-4mm}
\end{figure}

A simple model we use in this paper is shown in \emph{Fig.\ref{fig::network_model}}. The routing structure from the sender $f$ to the receivers $a$ and $b$ must be a tree rooted at $f$. Otherwise, there is routing loop which must be corrected. Assume that router $s$ is the ancestor node of both $a$ and $b$. Assume that the sender uses unicast to send messages to receivers, and assume that packets are sent in a back-to-back pair. For the \emph{k-}th pair of back-to-back packets, denoted as $a^k$ and $b^k$, sent from $f$ to $a$ and $b$, respectively, we use the following notation:

\begin{itemize}
  \item     $t_a(k)$: the time when $a$ receives $a^k$ in the \emph{k-}th pair.
  \item  	$t_b(k)$: the time when $b$ receives $b^k$ in the \emph{k-}th pair.
  \item 	$d_a(k)$: the latency of $a^k$ along the path from $f$ to $a$.
  \item 	$d_b(k)$: the latency of $b^k$ along the path from $f$ to $b$.
  \item  	$t_f(k)$: the time when $f$ sends the \emph{k-}th pair of packets.
\end{itemize}

\nop{The basic tree structure we use in this paper is shown in \emph{Fig.\ref{fig::network_model}}. Sender \emph{f} distributes a pair of back-to-back packets to two distinct receivers \emph{a}, \emph{b} respectively. Branching router \emph{s} is the \emph{parent/father} of \emph{a, b}: $s=f(a)$, $s=f(b)$. $t_a(k)$ denotes the time when the $k^{th}$ packet arrives at node \emph{a}. $t_b(k)$ at node \emph{b} is the same case while $t_f(k)$ at \emph{f} means the sending time of the $k^{th}$ packet pair at source \emph{f}. Let $d_a(k)$, $d_b(k)$ denote time latency on $path_a$, $path_b$ from \emph{f} to \emph{a}, \emph{b} in the $k^{th}$ transmission. $\delta$ indicates the time interval between one packet pair and the next. Delay correlation between nodes \emph{a}, \emph{b} is denoted as $\sigma_{a,b}^2$ or $\sigma_{d_a,d_b}^2$. In following paper we may use ``serial'' instead of ``pair'' for back-to-back packets if there are more than two receivers.}

Using network model of \emph{Fig.\ref{fig::network_model}} we obtain \emph{Eq.(\ref{eqn::t_a(0)=t_f(0)+d_a(0)})} for $a^0$ (we start the index with 0 for convenience):
\begin{equation}
\label{eqn::t_a(0)=t_f(0)+d_a(0)}
t_a(0)=t_f(0)+d_a(0)
\end{equation}

Similarly, for $a^k$ we have
\begin{equation}
  \label{eqn::t_a(k)=t_f(k)+d_a(k)}
  t_a(k)=t_f(k)+d_a(k)
\end{equation}

Let \emph{Eq.(\ref{eqn::t_a(k)=t_f(k)+d_a(k)})} $-$ \emph{Eq.(\ref{eqn::t_a(0)=t_f(0)+d_a(0)})} and $\delta_a(k)\equiv t_a(k)-t_a(0)$, we can obtain \emph{Eq.(\ref{eqn::a(k)delta =(t_f(k)-t_f(0))+(d_a(k)-d_a(0))})}:
\begin{equation}
  \label{eqn::a(k)delta =(t_f(k)-t_f(0))+(d_a(k)-d_a(0))}
  \delta_a(k) =(t_f(k)-t_f(0))+(d_a(k)-d_a(0))
\end{equation}

Denote the time interval between two consecutive pairs of packets as $\delta$. We assume that $\delta$ is a constant for simplicity at this moment, and relax this assumption later. In this case we use $k\delta\equiv k\cdot\delta$ to replace $t_f(k)-t_f(0)$ in \emph{Eq.(\ref{eqn::a(k)delta =(t_f(k)-t_f(0))+(d_a(k)-d_a(0))})}, then we have
\begin{equation}
  \label{eqn::d_a(k)=a(k)delta-kdelta+d_a(0)}
  d_a(k)=\delta_a(k)-k\delta+d_a(0)
\end{equation}

Let $\delta'_a(k)\equiv\delta_a(k)-k\delta$, \emph{Eq.(\ref{eqn::d_a(k)=a(k)delta-kdelta+d_a(0)})} is transformed into \emph{Eq. (\ref{eqn::d_a(k)=a(k)'delta+d_a(0)})}:
\begin{equation}
  \label{eqn::d_a(k)=a(k)'delta+d_a(0)}
  d_a(k)=\delta'_a(k)+d_a(0)
\end{equation}

We can achieve similar result at receiver $b$ as in \emph{Eq.(\ref{eqn::d_b(k)=b(k)'delta+d_b(0)})}:
\begin{equation}
  \label{eqn::d_b(k)=b(k)'delta+d_b(0)}
  d_b(k)=\delta'_b(k)+d_b(0)
\end{equation}
Where $\delta'_b(k)\equiv\delta_b(k)-k\delta$.

To estimate the correlation between $d_a(k)$ and $d_b(k)$, we introduce the following lemma.

\begin{Lemma}
  \label{Lemma::a,b,c,d}
  Assuming that $\zeta$, $\eta$ are two random variables, and $\chi=a\zeta+b$, $\gamma=c\eta+d$, where a,b,c,d are constant and a,c have the same symbol, thus we have $\sigma_{\zeta,\eta}^{2}=\sigma_{\chi,\gamma}^{2}$.
\end{Lemma}

\begin{IEEEproof}
\begin{equation}
  \begin{split}
    \sigma_{\chi,\gamma}^{2} & =\frac{E(\chi-E\chi)(\gamma-E\gamma)}{\sqrt{D\chi}\sqrt{D\gamma}}\\
    & =\frac{E(a\zeta+b-aE\zeta-b)(c\eta+d-cE\eta-d)}{\sqrt{a^2D\zeta}\sqrt{c^2D\eta}}\\
    & =\frac{acE(\zeta-E\zeta)(\eta-E\eta)}{|a||c|\sqrt{D\zeta}\sqrt{D\eta}}\\
    & =\frac{E(\zeta-E\zeta)(\eta-E\eta)}{\sqrt{D\zeta}\sqrt{D\eta}}\\
    & =\sigma_{\zeta,\eta}^{2}
  \end{split}
  \end{equation}
\end{IEEEproof}

Based on \emph{Lemma \ref{Lemma::a,b,c,d}}, we have the following theorem:
\begin{Theorem}
  \label{Therorem::correlation between a delta and b delta}
  The correlation between delay variables $d_a(k)$ and $d_b(k)$ is equivalent to the correlation between variables $\delta'_a(k)$ and $\delta'_b(k)$, which means
  \begin{equation}
    \sigma_{d_a(k),d_b(k)}^{2}=\sigma_{\delta'_a(k),\delta'_b(k)}^{2}
  \end{equation}
\end{Theorem}


Based on the measurements of $\delta'_a(k)$, $\delta'_b(k)$ we can calculate the correlation of delays along the path from $f$ to $a$ ($path_a$) and long the path from $f$ to $b$ ($path_b$), denoted as $\hat{\sigma}_{\delta'_a,\delta'_b}^{2}$ using \emph{Eq.(\ref{equ::a'delta,b'delta})}:

\begin{equation}
  \label{equ::a'delta,b'delta}
  \hat{\sigma}_{\delta'_a,\delta'_b}^{2}=\frac{1}{n-1}\sum_{k=1}^{n}[\delta'_a(k)-\overline{\delta'_a}][\delta'_b(k)-\overline{\delta'_b}]
\end{equation}
Where $\overline{\delta'_i}$ is the sample mean of ${\delta'_i(k)}_{k=1}^{n}$ for $i=a, b$.

\begin{Theorem}
  \label{Therorem::correlation unbiased estimator}
  $\hat{\sigma}_{\delta'_a,\delta'_b}^{2}$ in Eq.(\ref{equ::a'delta,b'delta}) is an unbiased estimator of the correlation on shared path.
\end{Theorem}

\begin{IEEEproof}
    First of all we show that $\hat{\sigma}_{d_a,d_b}^{2}$ (not $\hat{\sigma}_{\delta'_a,\delta'_b}^{2}$) is an unbiased estimator of the correlation on shared path $(f,s)$. Let $\lambda_a$, $\lambda_b$ denote the mean time latency of $path_a$, $path_b$ and let $\overline{d_a}$, $\overline{d_b}$ denote the sample mean correspondingly; true correlation ${\sigma}_{d_a,d_b}^{2}=E[(d_a(k)-\lambda_a)(d_b(k)-\lambda_b)]$. To prove that $E[\hat{\sigma}_{d_a,d_b}^{2}]=\sigma_{d_a,d_b}^{2}$ we analyze the expectation of $E[(d_a(k)-\overline{d_a})(d_b(k)-\overline{d_b}]$:
    \begin{equation}
     \label{equ::E[(d_a(k)-mean_a)(d_b(k)-mean_b)]}
     \begin{split}
       &E[(d_a(k)-\overline{d_a})(d_b(k)-\overline{d_b}]\\
       &=E[d_a(k)d_b(k)]-\frac{1}{n}\sum_{i=1}^{n}E[d_a(k)d_b(i)]\\
       &-\frac{1}{n}\sum_{i=1}^{n}E[d_a(i)d_b(k)]+\frac{1}{n^2}\sum_{i=1}^{n}\sum_{j=1}^{n}E[d_a(i)d_b(j)]
     \end{split}
    \end{equation}
    Since delays of the $i^{th}$ and $k^{th}$ pair are independent and
    \begin{equation*}
     \begin{split}
       {\sigma}_{d_a,d_b}^{2} & =E[(d_a(k)-\lambda_a)(d_b(k)-\lambda_b)]\\
       & =E[d_a(k)d_b(k)]-\lambda_a\lambda_b
     \end{split}
    \end{equation*}
    We obtain \emph{Eq.(\ref{equ::E[d_a(k)d_b(k)]})}
    \begin{equation}
     \label{equ::E[d_a(k)d_b(k)]}
     \begin{split}
       E[d_a(k)d_b(i)]= \left\{
           \begin{aligned}
           & \lambda_a\lambda_b     &k\neq i\\
           & \lambda_a\lambda_b+\sigma_{d_a,d_b}^{2}    &k=i\\
           \end{aligned}
        \right.
     \end{split}
    \end{equation}
    Substituting \emph{Eq.(\ref{equ::E[d_a(k)d_b(k)]})} into \emph{Eq.(\ref{equ::E[(d_a(k)-mean_a)(d_b(k)-mean_b)]})} we obtain
    \begin{equation*}
     \begin{split}
       E[(d_a(k)-\overline{d_a})(d_b(k)-\overline{d_b}]=(\frac{n-1}{n})\sigma_{d_a,d_b}^{2}
     \end{split}
    \end{equation*}
    Therefore, $\hat{\sigma}_{d_a,d_b}^{2}$ is an unbiased estimator of the correlation on shared path as is shown in \emph{Eq.(\ref{equ::dadb2})}
    \begin{equation}
     \label{equ::dadb2}
     \begin{split}
       E[\hat{\sigma}_{d_a,d_b}^{2}]& = \frac{1}{n-1}\sum_{k=1}^{n}E[(d_a(k)-\overline{d_a})(d_b(k)-\overline{d_b})]\\
       & = \frac{1}{n-1}n(\frac{n-1}{n})\sigma_{d_a,d_b}^{2}=\sigma_{d_a,d_b}^{2}
     \end{split}
    \end{equation}

    According to $\sigma_{d_a,d_b}^{2}=\sigma_{\delta'_a,\delta'_b}^{2}$ in \emph{Theorem \ref{Therorem::correlation between a delta and b delta}} we prove it.
\end{IEEEproof}

\vspace{-0.03cm}
\subsection{Discussion of $\delta$}
\vspace{-0.01cm}

\vspace{-0.03cm}
\subsubsection{$\delta$ is a constant}
\vspace{-0.01cm}

A complete delay correlation estimation algorithm (DCE) is summarized in \emph{Algorithm \ref{alg:DCEBased on PATI}} if the time interval $\delta$ is a constant.

\begin{algorithm}
\caption{Delay Correlation Estimation Algorithm}
\label{alg:DCEBased on PATI}
\begin{algorithmic}[1]
 \REQUIRE ~~\\
 Given time interval $\delta$.
 \ENSURE~~\\
 \FOR {$k=0:n$}
    \STATE Using \emph{Eq.(\ref{eqn::t_a(0)=t_f(0)+d_a(0)})} to \emph{Eq.(\ref{eqn::a(k)delta =(t_f(k)-t_f(0))+(d_a(k)-d_a(0))})} to measure $\delta_a(k)$ in the $k^{th}$ transmission;
    \STATE Using $\delta'_a(k)\equiv\delta_a(k)-k\delta$ to calculate $\delta'_a(k)$;
    \STATE Similarly measuring $\delta_b(k)$ and calculating $\delta'_b(k)\equiv\delta_b(k)-k\delta$;
 \ENDFOR
 \STATE For all the $\delta'_a(k)$ and $\delta'_b(k)$ using \emph{Eq.(\ref{equ::a'delta,b'delta})} to obtain the unbiased estimator for correlation $\sigma_{\delta'_a,\delta'_b}^{2}$, which is equivalent to $\sigma_{d_a,d_b}^{2}$ between \emph{a}, \emph{b} according to \emph{Theorem \ref{Therorem::correlation between a delta and b delta}}.
\end{algorithmic}
\end{algorithm}

\vspace{-0.03cm}
\subsubsection{$\delta$ is not a constant}
\vspace{-0.01cm}

If the time interval $\delta$ is not a constant then $t_f(k)-t_f(0)\neq k\delta$. In this case using $k\delta$ to replace $t_f(k)-t_f(0)$ is inappropriate and we choose $\delta_f(k)$ to denote $t_f(k)-t_f(0)$ in \emph{Eq.(\ref{eqn::a(k)delta =(t_f(k)-t_f(0))+(d_a(k)-d_a(0))})}, then \emph{Eq.(\ref{eqn::d_a(k)=a(k)delta-kdelta+d_a(0)})} can be rewritten to \emph{Eq.(\ref{equ::d_a(k)=delta_a(k)-delta_f(k)+d_a(0)})}.

\begin{equation}
  \label{equ::d_a(k)=delta_a(k)-delta_f(k)+d_a(0)}
  d_a(k)=\delta_a(k)-\delta_f(k)+d_a(0).
\end{equation}

Correspondingly, \emph{Eq.(\ref{eqn::d_a(k)=a(k)'delta+d_a(0)})} and \emph{Eq.(\ref{eqn::d_b(k)=b(k)'delta+d_b(0)})} can be replaced with $\delta'_a(k)\equiv\delta_a(k)-\delta_f(k)$ and $\delta'_b(k)\equiv\delta_b(k)-\delta_f(k)$, respectively.

Note that $t_f(k)$ is a timestamp contained in the packet, and thus $\delta_f(k)=t_f(k)-t_f(0)$ is readily available.

%

\vspace{-0.03cm}
\subsection{A mechanism for passive realization}
\vspace{-0.01cm}

To reduce explicit probing we propose a mechanism for passive realization of \emph{Algorithm \ref{alg:DCEBased on PATI}} in real networks.

\begin{figure}[htb]
\begin{center}
\epsfig{file=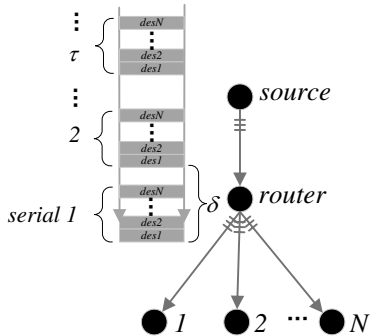, width=0.3\textwidth}
\vspace{-4mm}
\end{center}
\caption{\label{fig::passive_tomography_mechanism_realized} The mechanism for passive tomography with \emph{source} and \emph{N} end hosts.}
\vspace{-2mm}
\end{figure}


As \emph{Fig.\ref{fig::passive_tomography_mechanism_realized}} shows passive realization works as follows. In practical networks (for example, P2P networks) if \emph{N} end hosts request common contents from a \emph{source}, it will distribute packets. In this situation \emph{source} first chooses the \emph{No.1} requested data block which is duplicated into packet \emph{serial 1} and sent out to all \emph{N} hosts simultaneously guaranteeing that there exist two successive packets in a back-to-back manner. An indicator (\emph{\textbf{IR}}) is needed to tell if the received packet at each host belongs to the back-to-back pair. If serial of \emph{No.1} is sent completely \emph{source} repeats to the next until all requested contents are received by \emph{N} hosts. As regular data flow proceeds transmitting we change destination address of the current two successive packets when delay correlation between the corresponding host pair has been measured (if number of packets sent to them with indicator \emph{\textbf{IR}} reaches $\tau$ where $\tau$ is a tunable threshold).

One may naturally raise two questions: first is which two successive packets in one serial are chosen to add an \emph{\textbf{IR}}? while the other is how to guarantee that in each transmission the two successive packets are in a back-to-back manner? A simple mechanism can solve both of them. The basic idea is that we divide packets into small size. This can satisfy both the need of back-to-back and regular data transmission. In fact our experiment results show that any two successive packets in a serial distributed to \emph{N} hosts can be chosen to add \emph{\textbf{IR}} as long as the time interval $\delta$ is appropriate.

\vspace{-0.03cm}
\subsection{Arrangement of \emph{\textbf{IR}}}
\vspace{-0.01cm}

\begin{figure}[htb]
\begin{center}
\epsfig{file=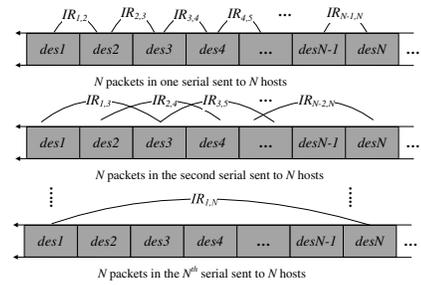, width=0.32\textwidth}
\vspace{-4mm}
\end{center}
\caption{\label{fig::arrange_IR} Arrangement of \emph{\textbf{IR}}. \emph{IR}$_{i,j}$ in each serial indicates blocks destined to \emph{i, j} should be successive.}
\vspace{-2mm}
\end{figure}

Based on above argument that each successive pair of packets can be regarded as back-to-back, in one transmitting serial at most \emph{N-1} delay correlations are measured (shown in \emph{Fig.\ref{fig::arrange_IR}}). After \emph{N-1} times switching of destination addresses and \emph{\textbf{IRs}} all delay correlations between \emph{N} hosts can be obtained.

In this way the complexity is only $O(N)$ compared with $O(N^2)$ to measure $N(N-1)/2$ correlations.

\vspace{-0.1cm}
\section{Simulation results}
\label{sec::experiments}
\vspace{-0.1cm}

\vspace{-0.02cm}
\subsection{Setups of simulation}
\vspace{-0.01cm}

\begin{figure}[htb]
\begin{center}
\epsfig{file=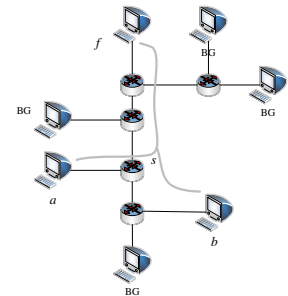, width=0.21\textwidth}
\vspace{-4mm}
\end{center}
\caption{\label{fig::simStru} Structure for DCE test}
\vspace{-4mm}
\end{figure}

We use OMNeT++ for simulations \cite{OMNet++::WebSite} to demonstrate the correctness and robustness of passive DCE tomography. We generate a network shown in \emph{Fig.\ref{fig::simStru}}. Nodes of \emph{BG} is for producing background traffic while others are the \emph{source} and client nodes. When two hosts $a$, $b$ request contents from $f$, it will send regular data in a back-to-back manner. In this case route algorithm determines a multicast tree with root $f$ and leaves $a$, $b$.

We set packet hundreds of bytes to satisfy both regular transmission and back-to-back property. One advantage is that since it is smaller than Maximum Transmission Unit (MTU) we avoid delay for package segmentation. We set bandwidth of each link value of 100Mbps; the background traffic pattern conforms to the Poisson distribution, whose expectation value could be set from 1MBps to 12MBps. We also change the size of packet from 100 bytes to MTU to see its influence on DCE measurement.

\vspace{-0.03cm}
\subsection{Results}
\vspace{-0.01cm}

Using DCE methodology we set $\tau$ to be 1550 and observe over 1500 timing samples for receiver pair.

\begin{figure}[htb]
\begin{center}
\epsfig{file=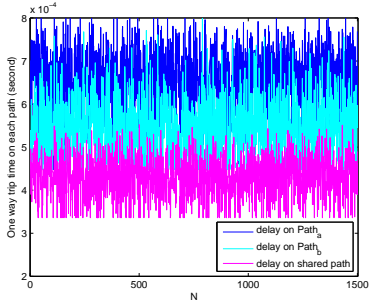, width=0.31\textwidth}
\vspace{-4mm}
\end{center}
\caption{\label{fig::delayValue_ab} One way trip time latency on $path_a$, $path_b$ and shared path $(f,s)$}
\vspace{-4mm}
\end{figure}

\emph{Fig.\ref{fig::delayValue_ab}} depicts the one way trip latency in our environment. One example of the average delay on $path_a$, $path_b$ and shared path are 0.6526ms, 0.5478ms and 0.4317ms respectively. In some case errors may happen to the timestamp as the variation of background traffic. Therefore, we ignore packets beyond twice the average delay on each path.

\begin{figure}[htb]
\begin{center}
\epsfig{file=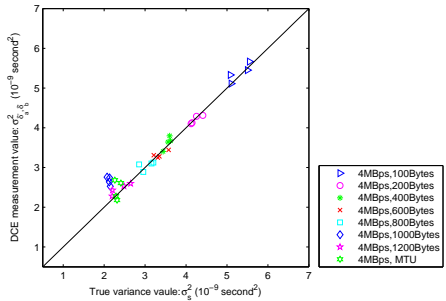, width=0.38\textwidth}
\vspace{-4mm}
\end{center}
\caption{\label{fig::cov_trueValue} DCE measurement covariance $\sigma_{\delta'_a,\delta'_b}^{2}$ versus the directly measured delay variance on the shared path $(f,s)$}
\vspace{-4mm}
\end{figure}

\emph{Fig.\ref{fig::cov_trueValue}} shows the DCE covariance $\sigma_{\delta'_a,\delta'_b}^{2}$ versus true value $\sigma_s^2$ on shared path $(f,s)$. Value $\sigma_{\delta'_a,\delta'_b}^{2}$ is calculated using the arriving time of package at $a$, $b$ while $\sigma_s^2$ is calculated directly from delay on $(f,s)$. According to \emph{Theorem \ref{Therorem::correlation between a delta and b delta}} we know that $\sigma_{d_a,d_b}^{2}$ is equal to $\sigma_{\delta'_a,\delta'_b}^{2}$ thus, ideally $\sigma_{\delta'_a,\delta'_b}^{2}$ and $\sigma_s^2$ should be identical and fall onto the 45 degree line. Taking test result with package size of 800 Bytes for example we see the estimated value is always locating nearby the true value with slight difference which demonstrates the correctness of DCE tomography. If we change size of data packet from 100 bytes to MTU delay correlation measured by DCE is always desired. This demonstrates that passive mechanism is able to achieve both data transmission and tomography.

\begin{figure}[htb]
\begin{center}
\epsfig{file=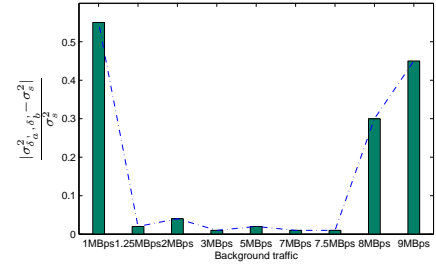, width=0.34\textwidth}
\vspace{-4mm}
\end{center}
\caption{\label{fig::cov_trueValueBG}$\frac{|\sigma_{\delta'_a,\delta'_b}^{2}-\sigma_s^2|}{\sigma_s^2}$: DCE measurement error versus different background traffic}
\vspace{-4mm}
\end{figure}

In \emph{Fig.\ref{fig::cov_trueValueBG}} packet size is fixed to be MTU. In this case performance of DCE tomography is perfect with the percentage error below 4\% when the expectation value of background traffic is within the region [1.25MBps,7.5MBps]. However, when it increases to 8MBps, error percentage increases sharply. This is because in this situation network's performance becomes worse and some shared paths between higher level routers are congested heavily, which destroys the back-to-back property of packets. Note that when the expectation value of background traffic is relatively small (below 1MBps) delay correlation caused by queuing on routers will not be significant thus the performance also degrades.

\vspace{-0.1cm}
\section{Conclusions and future work}
\label{sec::conclusions}
\vspace{-0.1cm}

In this paper we propose a novel tomography method named DCE to estimate delay correlation with no need of synchronization and cooperation between end hosts. We also develop the passive mechanism to further save bandwidth. Extensive simulations demonstrate the correctness of DCE. Moreover, passive realization is able to achieve both purpose of tomography and data transmission with excellent robustness versus different background traffic and package size.

In future, we plan to utilize the DCE measure for topology tomography.


\bibliographystyle{IEEEtran}
\bibliography{Reference}

\end{document}